\documentclass[a4paper]{spie}

\usepackage[]{graphicx}
\usepackage{aas_macros}
\usepackage[]{subfigure}

\title{An X-ray Polarimeter for HXMT Mission}

\author{Enrico~Costa\supit{a},  Ronaldo~Bellazzini\supit{c}, Gianpiero~Tagliaferri\supit{d}, Luca~Baldini\supit{c}, Stefano~Basso\supit{d}, Johan~Bregeon\supit{c}, Alessandro~Brez\supit{c}, Oberto~Citterio\supit{d}, Vincenzo~Cotroneo\supit{d}, Filippo~Frontera\supit{f,g}, Massimo~Frutti\supit{a}, Giorgio~Matt\supit{e}, Massimo~Minuti\supit{c}, Fabio~Muleri\supit{a,b}, Giovanni~Pareschi\supit{d}, Giuseppe~Cesare~Perola\supit{e}, Alda~Rubini\supit{a}, Carmelo~Sgro\supit{c}, Paolo~Soffitta\supit{a}, Gloria~Spandre\supit{c}
\skiplinehalf
\supit{a} Istituto di Astrofisica Spaziale e Fisica Cosmica, Via del Fosso del Cavaliere 100, I-00133 Roma, Italy;
\\
\supit{b} Universit\`{a} di Roma Tor Vergata, Dipartimento di Fisica, via della Ricerca Scientifica 1, 00133 Roma, Italy
\\
\supit{c} Istituto Nazionale di Fisica Nucleare, Largo B.
Pontecorvo 3, I-56127 Pisa,  Italy
\\
\supit{d} Osservatorio Astronomico di Brera, via E. Bianchi 46,
I-23807 Merate (Lc), Italy
\\
\supit{e} Universit\`{a} di Roma Tre, Dipartimento di Fisica, via
della Vasca Navale 84, I-00146 Roma , Italy
\\
\supit{f} Universit\`{a} di Ferrara, Dipartimento di Fisica, via
Saragat 1, I-44100 Ferrara, Italy
\\
\supit{g}Istituto di Astrofisica Spaziale e Fisica Cosmica, Via
Gobetti 101, I-40129 Bologna (BO), Italy}

\authorinfo{Further author information: (Send correspondence to Enrico Costa)
\\Enrico Costa: E-mail: enrico.costa@iasf-roma.inaf.it,
Telephone: +39-0649934004}

\begin{document}
\maketitle

\begin{abstract}
The development of micropixel gas detectors, capable to image
tracks produced in a gas by photoelectrons,  makes possible to
perform polarimetry of X-ray celestial sources in the focus of
grazing incidence X-ray telescopes.

HXMT is a mission by the Chinese Space Agency aimed to survey the
Hard X-ray Sky  with Phoswich detectors, by exploitation of the
direct demodulation technique. Since a fraction of the HXMT time
will be spent on dedicated pointing of particular sources, it
could host, with moderate additional resources a pair of X-ray
telescopes, each with a photoelectric X-ray polarimeter in the
focal plane.

We present the design of the telescopes and the focal plane
instrumentation and discuss the performance of this instrument to
detect the degree and angle of linear polarization of some
representative sources.

Notwithstanding the limited resources the proposed instrument can
represent a breakthrough in X-ray Polarimetry.
\end{abstract}


\section{Introduction}

According to an extended literature X-ray sources are expected to
show a significant degree of linear polarization. This can derive
from the non-thermal emission process itself or from the transfer
of the radiation in subsystems with geometries very far from the
spherical symmetry, such as accretion disks or columns. The
birefringence of plasma and of vacuum itself in the presence of
strong magnetic fields, stated by Quantum Electro Dynamics is the
source of an additional phenomenology, as it can, in different
conditions, polarize or depolarize the radiation or rotate its
polarization plane. As a matter of fact these physical conditions,
expected to produce polarization as a function of time or energy,
should be present in most of the astrophysical X-ray emitters. The
polarization should provide a tight test of any model far below
what required by spectra and variability alone. Significant
efforts were spent on polarimetry in the early phase of X-ray
Astronomy, until the late '70s. A series of rockets by the
Columbia University Team first detected the polarization from the
Crab Nebula\cite{Novick1972}. This result was improved by OSO-8
satellite\cite{Weisskopf1976}, that also derived upper limits on
polarization of other sources\cite{Silver1979,Long1980}. In the
following 30 years no more polarimeter has been launched aboard a
space mission. In fact the introduction of X-ray optics increased
the sensitivity of X-ray missions of orders of magnitude, in terms
of identification and imaging, while the sensitivity of
polarimeters, based on the bragg or scattering techniques, was
improving very slowly. Moreover X-ray telescopes abandoned the
requirement of rotating vehicle, while this was still a
requirement for polarimeters. In fact polarimetry was perceived as
low throughput subtopic, with sensitivity increasingly mismatched
with that of these new missions, targeted to different sources and
setting though requirements in terms of resources and operations.
A dedicated mission could be more suitable but has not been
approved so far. A new chance to rejuvenate this subtopics of
X-ray astronomy is yielded by the recent development of detectors
based on photoelectric effect in gas\cite{Bellazzini2007}. They
are truly imaging devices and, when combined with X-ray telescopes
permit to extend to polarimetry the same huge increase of
sensitivity, achieved so far for imaging and spectroscopy. They
also allow for a good timing and a resolution suitable for energy
resolved polarimetry of continua. In any case, starting from the
data of OSO-8 and taking into account the modern theoretical
analysis polarizations to be typically expected from celestial
sources are in the range of 1 to a few \%. In statistical terms
this means that sources are to be observed till photons are
detected in numbers of the order of $10^{5} - 10^{6}$. This
implies that polarimetry is still confined to a limited number of
brighter sources. Therefore three type of mission scenarios can be
foreseen: a dedicated mission, a mission with one telescope and
different instruments to be alternatively set in the focus, a
mission with different instruments in parallel with a reasonable
overlap of scientific objectives.

A certain connection does exist between physics of hard X-ray
emitters and expectations of polarization. Non thermal processes,
can be singled out by the presence of \textit{hard tails} as well
as from the existence of linear polarization. The latter also
provides a geometric information (e.g. the orientation of magnetic
fields or the direction of particle acceleration). Alternatively
the presence of hard component and the absence of polarization can
provide the evidence for disordered systems (e.g. disordered
magnetic fields for synchrotron, or diffuse source of seed photons
in an inverse compton). Because of these overlaps and because of a
reduced mismatching of observing times, a polarimeter and a hard
X-ray instrument can efficiently combine, resulting in a very
performing mission for deep Physics of non thermal sources. The
Hard X-ray Modulation Telescope mission HXMT\cite{LiTP2006}, based
on an array of phoswich detectors with slat collimators, is mainly
devoted to performing a hard X-ray all-sky imaging survey with
both high sensitivity, spatial of 5 arcminutes and positioning
accuracy of 1arcminute. HXMT can also make high signal-to-noise
ratio pointing observations of sources of particular scientific
interest. HXMT ia now in the full design phase and a candidate for
the first Chinese dedicated astronomy satellite. Since a
combination of Hard X-ray measurements with polarimetry seems to
be very promising, Italian and Chinese Space Agencies are
negotiating the possible inclusion in the same bus of two X-ray
telescopes, with in the focus two gas pixel polarimeters.

\section{The Telescopes} \label{sec:Telescopes}

The design of the proposed telescopes is based on the following
concepts:
\begin{itemize}
\item{be compliant with the allocated weight and volume}
\item{be based on solid technologies with some limited improvements for the particular application}
\item{relax some features that cannot be coped by the detector in favor of a larger collecting area }
\end{itemize}

Since the schedule of the Mission is relatively ambitious we
selected the technology of producing the telescope shells by
replicating superpolished mandrels with electroforming. This
technology, developed for SAX\cite{Conti1994} and successfully
applied to XMM\cite{Gondoin1994} and JET-X\cite{Citterio1995}
(whose spare unit is the optics of SWIFT X-Ray
Telescope\cite{Burrows2000}), is adequately under control for a
short track mission. A major improvement is the use of Iridium as
a reflector. The constraint of a maximum length of 2.5 meters has
fixed the mirror focal length to 2.1 meters. The Ir provides an
improvement at higher energies. Another improvement of high impact
is the addition fa thin Carbonium coating. In fact, taking into
account the dependence on the energy of the modulation factor,  of
the reflectivity of materials and the steepness of the spectra of
sources, the sensitivity of the polarimeter is maximal around 2-4
keV. In this range the M edges of Ir (but Au would not be better)
introduce a strong decrease of reflectivity. It has been proofed
that a thin Carbonium coating can significantly reduce the
effect\cite{Pareschi2004}. This treatment is very important in
general, but is particularly effective if the telescope is used
for polarimetry.

\begin{figure}[htbp]
\begin{center}
\includegraphics[angle=0,width=16cm]{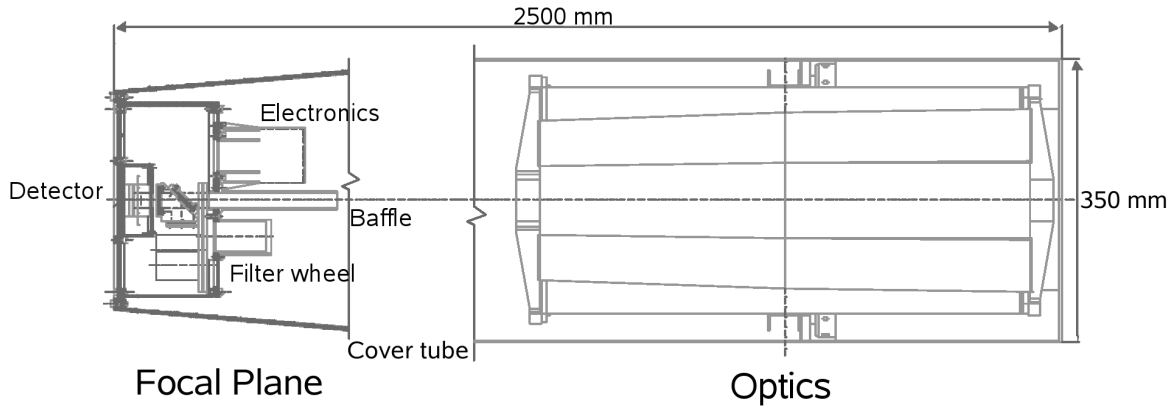}
\caption{The telescope} \label{telescope}
\end{center}
\end{figure}

The gas pixel detector has an intrinsic resolution of the order of
150 $\mu$m, but in the focus of a telescope photons impinge
inclined on the gas cell. With an absorption gap of 10 mm the
uncertainty on the depth of the interaction results in a further
blurring of the order of 400  $\mu$m. For this relatively short
focal length this additional effect limits the angular resolution
to $\sim$ 40 arc seconds. For this reason and thanks to an
improved design to reduce the tension at the interface with the
mounting structure, we assume that a thickness of 100 $\mu$m for
the shells is sufficient. The total weight of the 30 shells is
10.5 kg. In fig.~ \ref{telescope} we show the drawing of the
mirror package, the focal plane and the carbon fibre supporting
structure. Shells are mounted, as usual, with spiders. Since the
detector is sensitive only above 1.5 keV we can afford a thermal
shield in the front and can avoid the heavy and long
collimator/buffle in front of the mirrors.

\begin{figure}[htbp]
\begin{center}
\includegraphics[angle=0,totalheight=10cm]{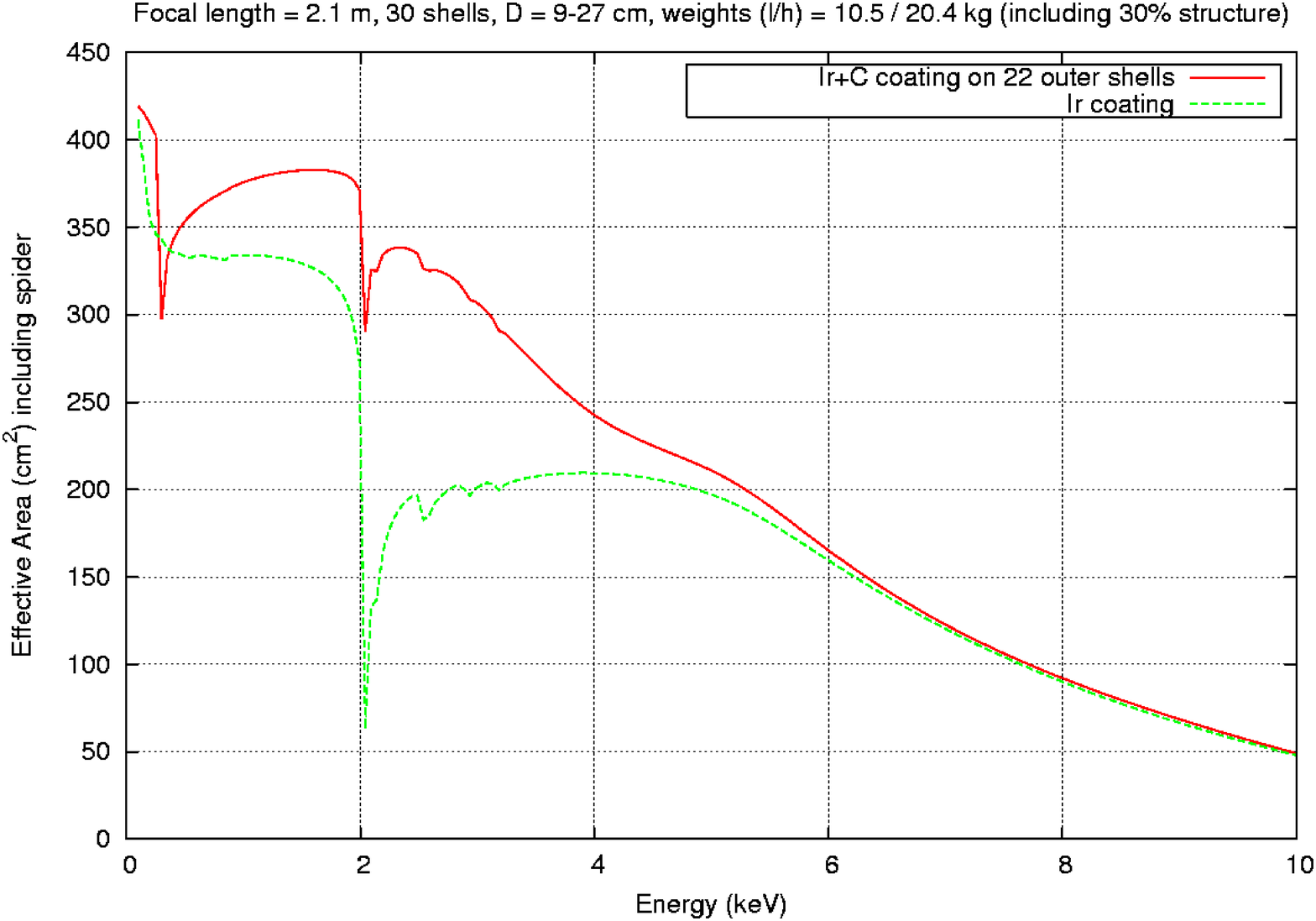}
\caption{The effective area of one telescope} \label{Effarea}
\end{center}
\end{figure}

In fig.~ \ref{Effarea} we show the effective area of one
telescope, with and without the coating of carbon. The
introduction of the latter is very effective indeed. The reduction
due to the spider is already accounted.

\begin{table}[htbp]
\begin{center}
\caption{Features of Each Telescope} \label{tab:TabTel}
\begin{tabular}{r|c}
\hline
Shell Design         &  Wolter 1                        \\
Shell Length         &  30 cm $\times$ 2                \\
Focal Length         &  2.1 m                           \\
Number of Shells     &  30                              \\
Shell Thickness      &  100 $\mu$m                      \\
Coating              &  Ir + C (only 22external)        \\
Mounting             &  spider                          \\
Weight of shells     &  8.1 kg                          \\
Total Weight         &  41.4 kg                         \\
\end{tabular}
\end{center}
\end{table}
%
%
%
\section{The Focal Plane} \label{sec:Focalplane}
\begin{figure}[htbp]
\begin{center}
\includegraphics[angle=0,totalheight=7cm]{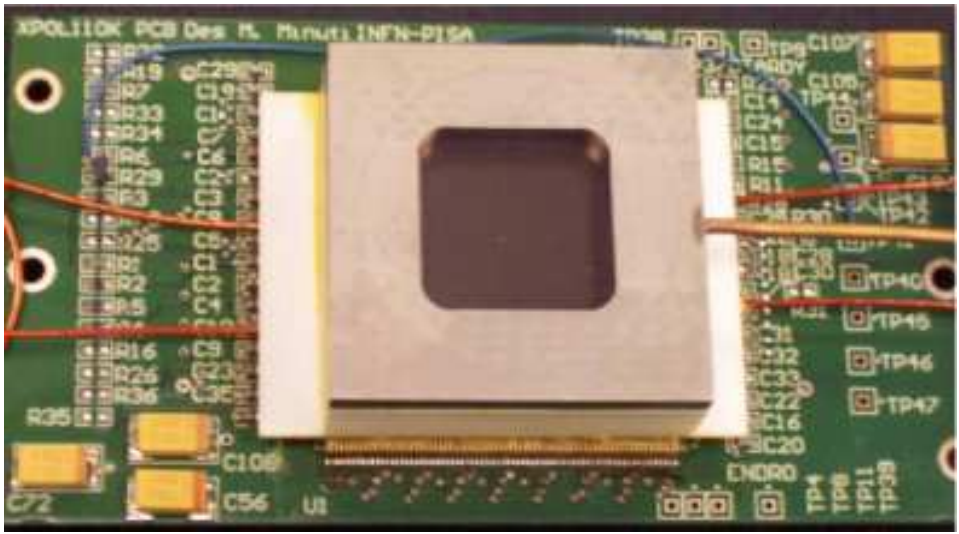}
\caption{A prototype of sealed micropixel detector}
\label{Detector}
\end{center}
\end{figure}

\begin{figure}[htbp]
\begin{center}
\includegraphics[angle=0,totalheight=5cm]{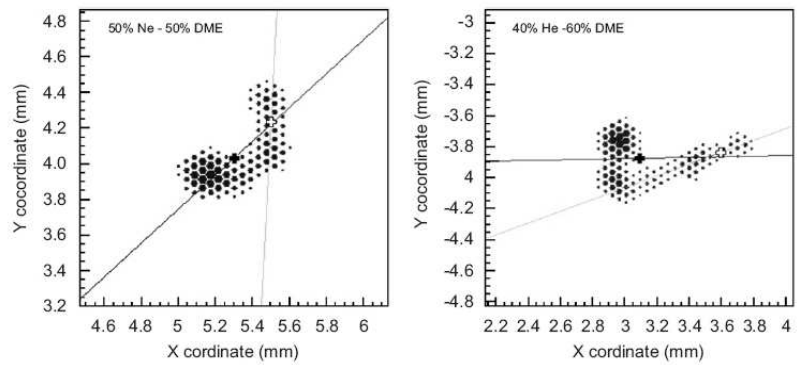}
\caption{Two examples of photoelectron tracks as visualized by the
micropixel detector} \label{Track}
\end{center}
\end{figure}

In the focus of each telescope there is a gas pixel detector. This
device has been developed by the Pisa INFN team. It is based on
the imaging of the tracks of photoelectrons created by conversion
of X-ray photons in a low Z gas mixture. The electrons of the
track are drifted by an electric field to a Gas Electron
multiplier that multiplies them in a proportional manner, while
preserving the shape of the track. The amplified electrons are
collected by metal pads, close to the GEM, that are on the top
layer of a VLSI chip, that includes, below the projection of the
pad, a complete and independent electronic chain. From the
analysis of the charge collected on each pad an image of the track
is obtained and the interaction point of the photon and the
initial direction of the photoelectron are estimated. The most
recent implementation of the VLSI chip has $\sim$ 100000 pixel
with a pitch of 50 $\mu$m on an hexagonal
pattern\cite{Bellazzini2006}. The chip has self-triggering
capability and fetches to the output only the content of a Region
of Interest around pixels that triggered.  Sealed prototypes of
it, with a Beryllium window 50 $\mu$m thick, have been recently
developed and manufactured a nd show a good stability versus time
and radiation\cite{Bellazzini2006c}. In fig.~ \ref{Detector} we
show a prototype of sealed detectors. In fig.~ \ref{Track} we show
two examples of tracks produced on the detector by photons from an
X-ray generator with Cr anode. Various mixtures have been tested.
The most performing with a band pass like that of HXMT telescope
is presently a mixture of He (20$\%$) and DME (80$\%$), which is
the current baseline.

\begin{figure}[htbp]
\begin{center}
\includegraphics[angle=0,totalheight=10cm]{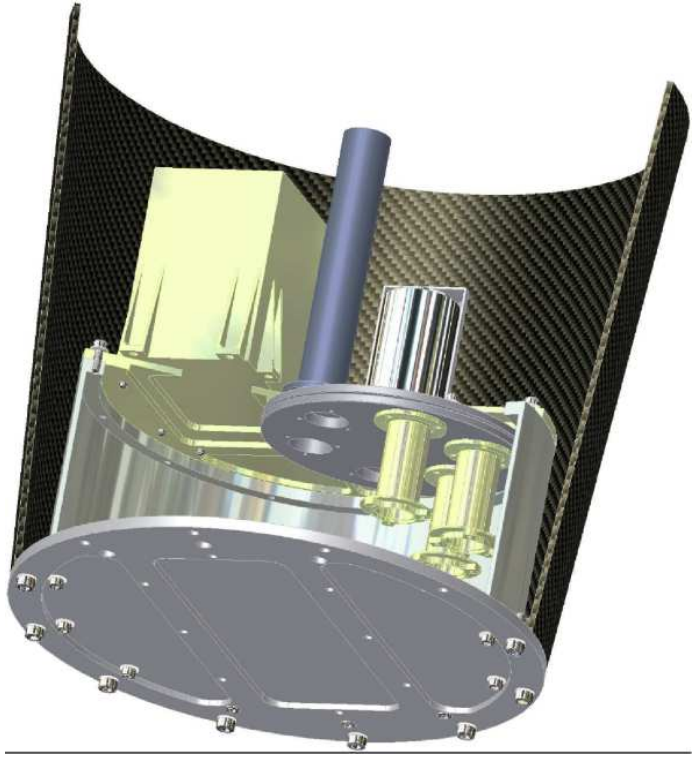}
\caption{The focal plane} \label{Assieme}
\end{center}
\end{figure}

The detector is connected with a flexible cable to an interface
electronics performing the A/D conversion and routing the controls
to the VLSI chip. Beside the detector and the interface
electronics the focal plane includes a filter wheel to allow to
cover the window, for protection, or to position in front of it a
polarized\cite{Muleri2007} or an unpolarized calibration source.
Also a long buffle has been included to prevent stray light and
ions that could arrive to the window. The whole is shown in fig.~
\ref{Assieme}.

The control electronics including High Voltage Power Supplies,
DC-DC converters, PDHU, Mass Memory, houskeeping and interface to
telemetry, is located separately from the telescope.

While the requirements for the detector and the electronics are
very moderate in terms of power and weight a large amount of
information must be transmitted. In fact the excellent noise
figure of the ASIC chip (50 electrons ENC\cite{Bellazzini2006}),
allows, with a gain of $\sim$ 500 from the GEM, for the detection
of single electrons generated in the gas. This, combined with the
pitch of 50 $\mu$m for the pixels, gives the capability to
preserve very detailed information on the track, but requires an
adequate amount of bit. Photons of 2 keV generate tracks with, in
average, 40 pixels with charge content above zero. This number
increases to 87 for photons of 6 keV. With the foreseen telescopes
and detectors we expect from the Crab 266 counts/s, corresponding
to around 250 kbit/s, after zero suppression. This number can be
reduced by means of various compression techniques. The most
radical would be an analysis of the track onboard, that would
reduce the rate to $\sim$ 16 kbit/s.

\section{The expected performances} \label{sec:Performances}
In table~\ref{TabGen} we resume the main overall features of the
proposed instruments
\begin{table}[htbp]
\begin{center}
\caption{Overall Features of the proposed instrument (two
telescopes)} \label{TabGen}
\begin{tabular}{r|c}
\hline
Effective Area @ 3 keV         &  740 cm $^{2}$      \\
f.o.v.        &  22' $\times$ 22'                \\
Window         &  50$\mu$m Beryllium                    \\
Gas filling     &  DME (80$\%$) He (20$\%$)             \\
Gas Pressure      &  1 atm                      \\
Absorption gap  &  10 mm                        \\
Angular Resolution       &  1'                        \\
Energy Resolution     &  $\frac{\Delta E}{E} = 0.2 \times (6/E)^{0.5}$                       \\
Energy Band       &  1.5 - 10 keV                \\
Timing       &  10 $\mu$s                         \\
\end{tabular}
\end{center}
\end{table}
\subsection{The statistics}

The sensitivity to sources is usually expressed as:
\begin{equation}\label{MDP}
  MDP=\frac{4.29}{\mu\epsilon F}\times\sqrt{\frac{B+\varepsilon F}{ST}}
\end{equation}

\begin{figure}[htbp]
\begin{center}
\includegraphics[angle=90,totalheight=10cm]{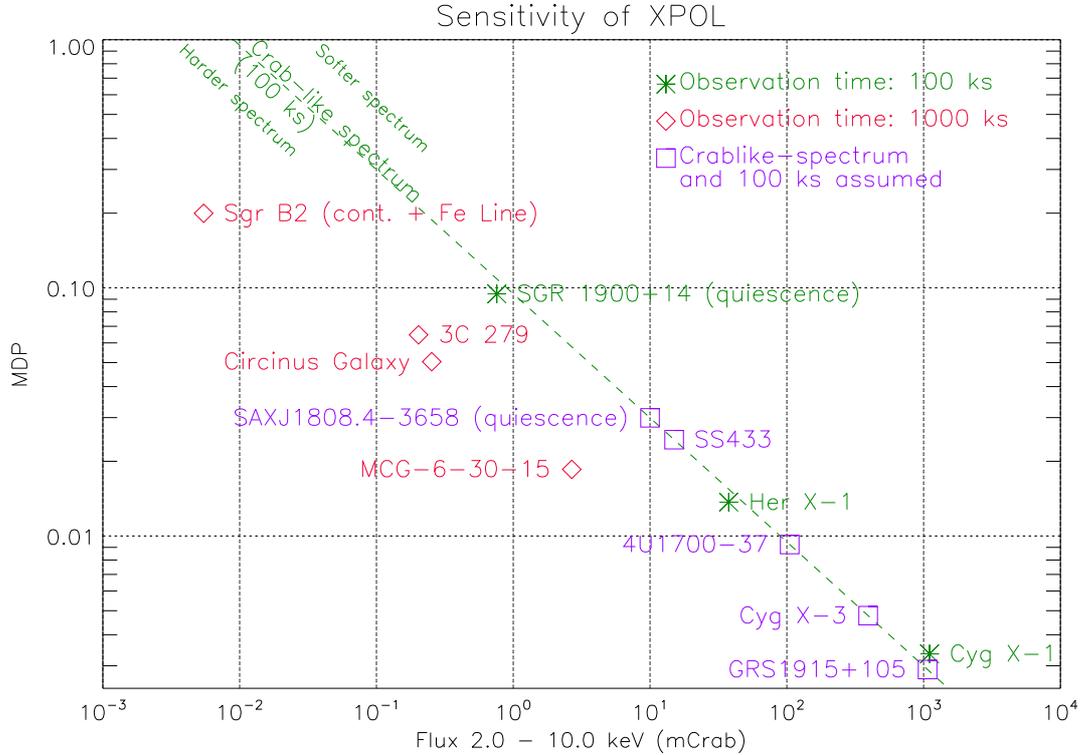}
\caption{The Minimum Detectable Polarization for observations of
$10^{5}$ or $10^{6}$ seconds with HXMT} \label{HXMTSensitivity}
\end{center}
\end{figure}

Where MDP is the Minimum Detectable Polarization at 99$\%$
confidence, $\mu$ is the modulation factor, $\epsilon$ is the
efficiency, S is the flux from the source, B is the background per
unit of area and T is the observing time. In the case of HXMT the
background is negligible to any practical effect. The modulation
factor has been computed by Monte Carlo simulations, which have
been recently confirmed by laboratory testing with polarized,
monochromatic sources at energies of 2.6, 3.7 and 5.2
KeV\cite{Muleri2007b} . We can therefore compute MDP for some
sources peculiar or representative of various classes. The results
are shown in fig.~ \ref{HXMTSensitivity}.

One day of observation can be sufficient to study bright galactic
sources, such as Cyg-X3, GRS1915+105 or 4U1700-37. With a few days
of observation phase resolved polarimetry of Her X-1 or Vela X-1
could be performed to a few $\%$ level, allowing for a direct
measurement of the inclination of the magnetic axis with respect
to the rotation axis and to the sky: a breakthrough for the
detailed modelling of these systems. A few bright extragalactic
sources could be detected with one week pointing. Also relatively
faint sources like Circinus Galaxy or 3C279 could be observed,
since high degrees of polarization can be expected. But in the
following we want to propose two cases of particular interest.

\subsection{The sensitivity to the angle: the case Cyg-X1}
\begin{figure}[htbp]
\begin{center}
\includegraphics[angle=90,totalheight=10cm]{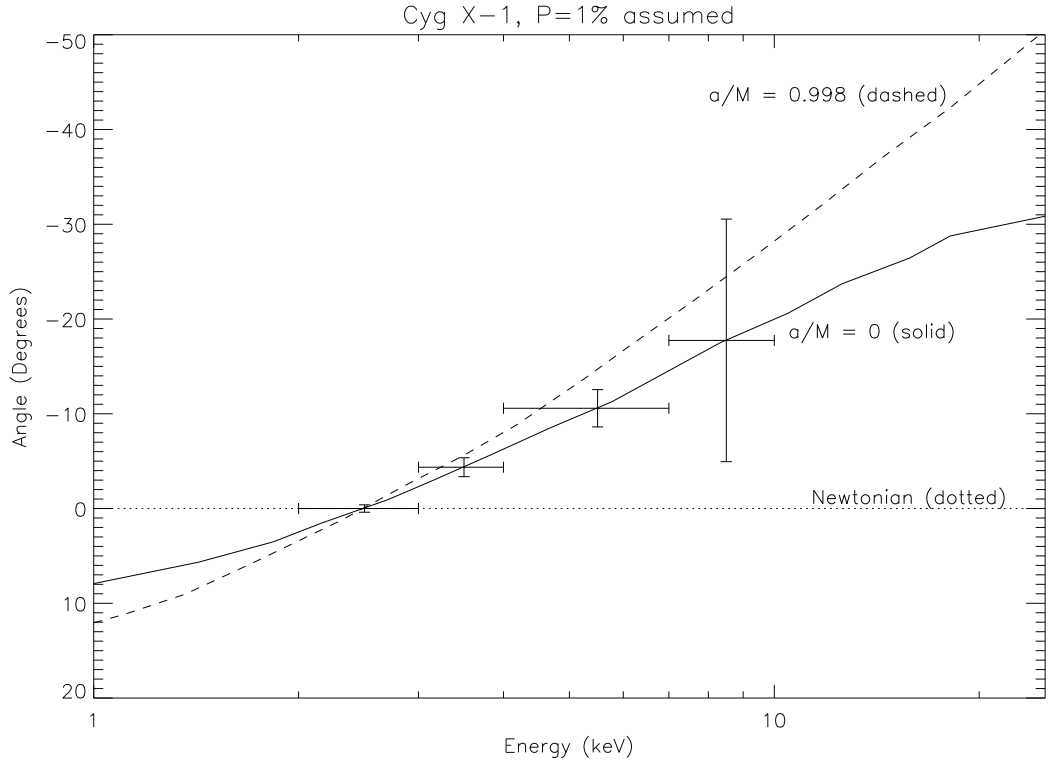}
\caption{The Rotation of the polarization angle for Cyg-X1. The
two curves represent the two extreme cases of a Schwarzschild BH
and a Kerr BH} \label{CygX1}
\end{center}
\end{figure}

In an accretion disk the polarization, produced by scattering on
selected directions, will be always perpendicular or parallel to
the disk. In the low energy case Chandrasekhar\cite{Chandra1950}
has demonstrated that polarization cannot exceed the limit of 11.7
\%. Sunyaev and Thitarchiuck\cite{SunTit1985} have demonstrated
that in the X-ray band, when the energy of the scattered photons
can be higher than that of the source photons, due to inverse
Compton, the polarization can significantly exceed the
Chandrasekhar limit. But, as outlined by Connors, Stark and
Piran\cite{Connors1980}, another effect will be present as well.
The X rays will be parallel or perpendicular to the disk in the
rest frame, but in the travel to the observer, will experience the
strong gravitational field of the Black Hole, that in the observer
frame will result in a rotation of the polarization plane. Since
the photons of different energies will derive from regions of the
disk at different distance from the BH, the total result will be a
rotation of the polarization plane as a continuous function of the
energy, a unique signature of the presence of a Black Hole. In
fig.~ \ref{CygX1} we show the capability of HXMT to detect this
effect if a polarization of 1$\%$ is there. This is conservative.
If the polarization is higher, as suggested by a marginal
detection with OSO-8\cite{Long1980} the polarimeter will allow for
the discrimination between a Schwarzschild BH and a Kerr BH.

\subsection{The sensitivity to the angle: the case of Sgr B2}

\begin{figure}[htbp]
\begin{center}
\subfigure[\label{Comptonangle}]{\includegraphics[angle=0,totalheight=4cm]{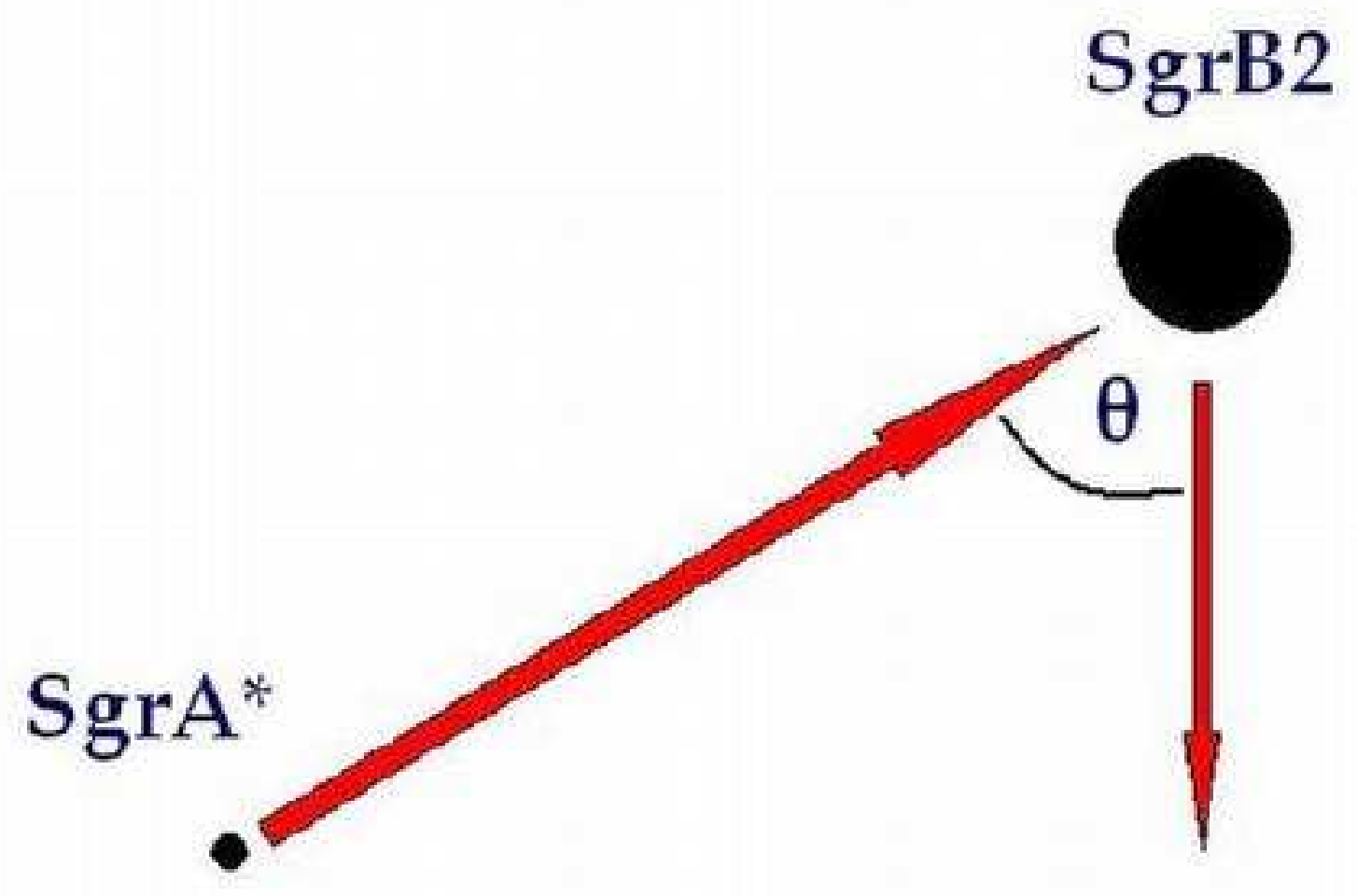}}
\subfigure[\label{IBISSgr}]{\includegraphics[angle=0,totalheight=5cm]{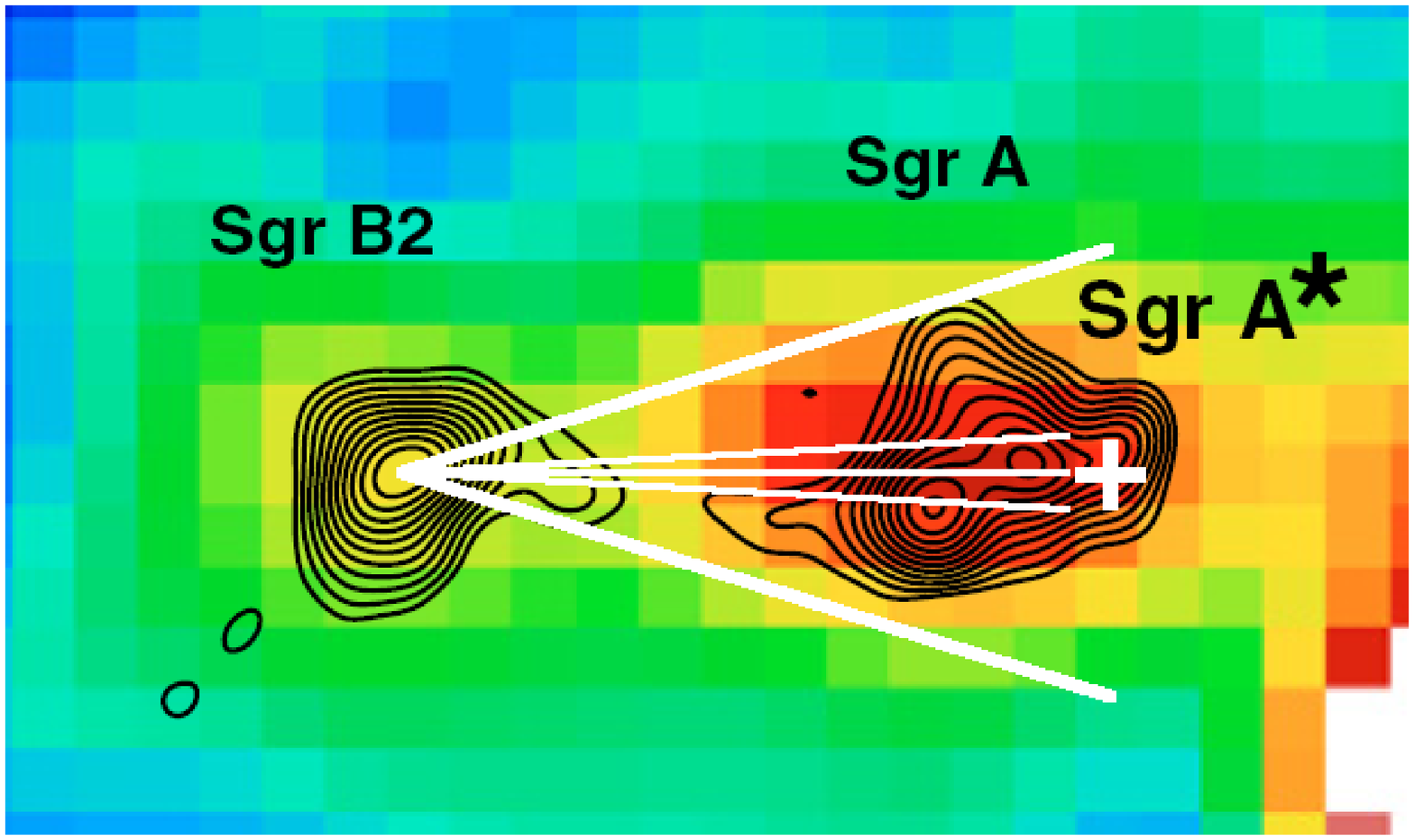}}
\end{center}
\caption{If the X-rays from SgrB2 are coming from SgrA$\*$ and are
reflected toward us, the amount of polarization will depend on the
angle $\vartheta$. From the measurement we can determine
$\vartheta$ and thence the distance to us. If the hypothesis is
correct the cone of confidence will include the source of the
scattered photons and make the association between the two sources
much stronger than from spectral arguments. The precision on the
angle will depend on the degree of polarization. We superimpose to
the IBIS map\cite{Revnivtsev2004} the cases of 70$\%$ and 10$\%$
polarization}
\end{figure}

The center of our Galaxy harbors a 2.6$\times10^6$ solar masses
Black Hole. The Black Hole is very quiet, its accretion luminosity
being about 10 orders of magnitude lower than the Eddington
luminosity. This inactivity is shared by most of supermassive BH
at the center of galaxies. only a small fraction are very active
and there is evidence of turning from one state to the other. But
at the projected distance of about 100 pc from the Black Hole,
there is a giant molecular cloud, Sgr B2(Fig.~ \ref{IBISSgr}),
which in X-rays has a pure reflection spectrum\cite{Koyama1996,
Revnivtsev2004 }). However, it is not clear what Sgr B2 is
reflecting: there are no bright enough sources in the vicinity.
The simplest explanation is that a few hundreds years ago our own
Galactic Center was much brighter, at the level of a low
luminosity Active Galaxy: the molecular cloud would then simply
echoing the past activity (\cite{Sunyaev1993}, \cite{Koyama1996}).
If this is true, as shown in Fig.~ \ref{Comptonangle}, the
reflected X-rays should be highly polarized: the degree will
depend on the angle $\vartheta$ and for $\theta$=0 will be of
100$\%$ for the continuum (the fluorescence will be unpolarized in
any case). From the polarization we can derive $\theta$ and from
that the distance of the source and the time when our own galaxy
was a little AGN\cite{Churazov2002}. Moreover the perpendicular to
the polarization plane will \textit{point} to the source of the
reflected photons. For a polarization of $\geq$70$\%$ (that
corresponds to angles from 60 to 120 $^{o}$ the error on the angle
will be lower than 3$^{o}$ and the association of the Sgr B2 with
the BH will be very cogent as shown in fig. ~ \ref{IBISSgr}. Of
course if $\vartheta$ is far from 90$^{o}$ the error would be
larger.
%

\section{Conclusion} \label{sec:Conclusion}

The two telescopes proposed for HXMT, based on established
technologies, can perform polarimetry on several classes of
objects and would boost this subtopics of X-ray astronomy, so far
exploited only marginally.

\section*{Acknowledgments}

This research is supported by ASI contract I/088/06/0.

\bibliography{ReferencesHX}   

\begin{thebibliography}{10}

\bibitem{Novick1972}
R.~{Novick}, M.~C. {Weisskopf}, R.~{Berthelsdorf}, R.~{Linke}, and R.~S.
  {Wolff}, ``{Detection of X-Ray Polarization of the Crab Nebula},'' {\em
  \apjl}~{\bf 174}, pp.~L1+, May 1972.

\bibitem{Weisskopf1976}
M.~C. {Weisskopf}, G.~G. {Cohen}, H.~L. {Kestenbaum}, K.~S. {Long},
  R.~{Novick}, and R.~S. {Wolff}, ``{Measurement of the X-ray polarization of
  the Crab Nebula},'' {\em \apjl}~{\bf 208}, pp.~L125--L128, Sept. 1976.

\bibitem{Silver1979}
E.~H. {Silver}, M.~C. {Weisskopf}, H.~L. {Kestenbaum}, K.~S. {Long},
  R.~{Novick}, and R.~S. {Wolff}, ``{The first search for X-ray polarization in
  the Centaurus X-3 and Hercules X-1 pulsars},'' {\em \apj}~{\bf 232},
  pp.~248--254, Aug. 1979.

\bibitem{Long1980}
K.~S. {Long}, G.~A. {Chanan}, and R.~{Novick}, ``{The X-ray polarization of the
  Cygnus sources},'' {\em \apj}~{\bf 238}, pp.~710--716, June 1980.

\bibitem{Bellazzini2007}
R.~{Bellazzini}, G.~{Spandre}, M.~{Minuti}, L.~{Baldini}, A.~{Brez},
  F.~{Cavalca}, L.~{Latronico}, N.~{Omodei}, M.~{Razzano}, F.~{Angelini}, M.~M.
  {Massai}, C.~{Sgr{\'o}}, E.~{Costa}, and P.~{Soffitta}, ``{X-ray polarimetry
  with Gas Pixel Detectors: A new window on the X-ray sky},'' {\em Nuclear
  Instruments and Methods in Physics Research A}~{\bf 576}, pp.~183--190, June
  2007.

\bibitem{LiTP2006}
T.~P. {Li}, S.~N. {Zhang}, H.~Y. {Wang}, L.~H. {Jiang}, M.~{Wu}, F.~J. {Lu},
  J.~M. {Wang}, L.~M. {Song}, B.~B. {Wu}, and Y.~{Chen}, ``{HXMT - A Chinese
  high energy astrophysics mission},'' in {\em 36th COSPAR Scientific
  Assembly},  {\em COSPAR, Plenary Meeting} {\bf 36}, pp.~2815--+, 2006.

\bibitem{Conti1994}
G.~{Conti}, E.~{Mattaini}, E.~{Santambrogio}, B.~{Sacco}, G.~{Cusumano},
  O.~{Citterio}, H.~W. {Braeuninger}, and W.~{Burkert}, ``{Engineering
  qualification model of the SAX x-ray mirror unit: technical data and x-ray
  imaging characteristics},'' in {\em Proc. SPIE Vol. 2011, p. 118-127,
  Multilayer and Grazing Incidence X-Ray/EUV Optics II, Richard B. Hoover;
  Arthur B. Walker; Eds.},  R.~B. {Hoover} and A.~B. {Walker}, eds., {\em
  Presented at the Society of Photo-Optical Instrumentation Engineers (SPIE)
  Conference} {\bf 2011}, pp.~118--127, Feb. 1994.

\bibitem{Gondoin1994}
P.~{Gondoin}, K.~{van Katwijk}, B.~R. {Aschenbach}, N.~{Schulz}, R.~{Boerret},
  H.~{Glatzel}, and O.~{Citterio}, ``{X-ray spectroscopy mission (XMM)
  telescope development},'' in {\em Proc. SPIE Vol. 2209, p. 438-450, Space
  Optics 1994: Earth Observation and Astronomy, M. G. Cerutti-Maori; Philippe
  Roussel; Eds.},  M.~G. {Cerutti-Maori} and P.~{Roussel}, eds., {\em Presented
  at the Society of Photo-Optical Instrumentation Engineers (SPIE) Conference}
  {\bf 2209}, pp.~438--450, Sept. 1994.

\bibitem{Citterio1995}
O.~{Citterio}, P.~{Conconi}, M.~{Ghigo}, F.~{Mazzoleni}, E.~{Poretti},
  G.~{Conti}, G.~{Cusumano}, B.~{Sacco}, H.~W. {Braeuninger}, and W.~{Burkert},
  ``{Status of the qualification model of the x-ray optics for the JET-X
  telescope aboard the Spectrum X-Gamma satellite},'' in {\em Proc. SPIE Vol.
  2515, p. 44-54, X-Ray and Extreme Ultraviolet Optics, Richard B. Hoover;
  Arthur B. Walker; Eds.},  R.~B. {Hoover} and A.~B. {Walker}, eds., {\em
  Presented at the Society of Photo-Optical Instrumentation Engineers (SPIE)
  Conference} {\bf 2515}, pp.~44--54, June 1995.

\bibitem{Burrows2000}
D.~N. {Burrows}, J.~E. {Hill}, J.~A. {Nousek}, A.~A. {Wells}, A.~D. {Short},
  R.~{Willingale}, O.~{Citterio}, G.~{Chincarini}, and G.~{Tagliaferri},
  ``{Swift X-Ray Telescope},'' in {\em Proc. SPIE Vol. 4140, p. 64-75, X-Ray
  and Gamma-Ray Instrumentation for Astronomy XI, Kathryn A. Flanagan; Oswald
  H. Siegmund; Eds.},  K.~A. {Flanagan} and O.~H. {Siegmund}, eds., {\em
  Presented at the Society of Photo-Optical Instrumentation Engineers (SPIE)
  Conference} {\bf 4140}, pp.~64--75, Dec. 2000.

\bibitem{Pareschi2004}
G.~{Pareschi}, V.~{Cotroneo}, D.~{Spiga}, D.~{Vernani}, M.~{Barbera}, M.~A.
  {Artale}, A.~{Collura}, S.~{Varisco}, G.~{Grisoni}, G.~{Valsecchi}, and
  B.~{Negri}, ``{Astronomical soft x-ray mirrors reflectivity enhancement by
  multilayer coatings with carbon overcoating},'' in {\em UV and Gamma-Ray
  Space Telescope Systems. Edited by Hasinger, G{\"u}nther; Turner, Martin J.
  L. Proceedings of the SPIE, Volume 5488, pp. 481-491 (2004).},  G.~{Hasinger}
  and M.~J.~L. {Turner}, eds., {\em Presented at the Society of Photo-Optical
  Instrumentation Engineers (SPIE) Conference} {\bf 5488}, pp.~481--491, Oct.
  2004.

\bibitem{Bellazzini2006}
R.~{Bellazzini}, G.~{Spandre}, M.~{Minuti}, L.~{Baldini}, A.~{Brez},
  F.~{Cavalca}, L.~{Latronico}, N.~{Omodei}, M.~M. {Massai}, C.~{Sgro'},
  E.~{Costa}, P.~{Soffitta}, F.~{Krummenacher}, and R.~{de Oliveira}, ``{Direct
  reading of charge multipliers with a self-triggering CMOS analog chip with
  105 k pixels at 50 {$\mu$}m pitch},'' {\em Nuclear Instruments and Methods in
  Physics Research A}~{\bf 566}, pp.~552--562, Oct. 2006.

\bibitem{Bellazzini2006c}
R.~{Bellazzini}, G.~{Spandre}, M.~{Minuti}, L.~{Baldini}, A.~{Brez},
  L.~{Latronico}, N.~{Omodei}, M.~{Razzano}, M.~M. {Massai}, M.~{Pinchera},
  M.~{Pesce-Rollins}, C.~{Sgro}, E.~{Costa}, P.~{Soffitta}, H.~{Sipila}, and
  E.~{Lempinen}, ``{A Sealed Gas Pixel Detector for X-ray Astronomy},'' {\em
  ArXiv Astrophysics e-prints} , Nov. 2006.

\bibitem{Muleri2007}
F.~{Muleri}, P.~{Soffitta}, R.~{Bellazzini}, A.~{Brez}, E.~{Costa},
  S.~{Fabiani}, M.~{Frutti}, M.~{Minuti}, M.~B. {Negri}, P.~{Pascale},
  A.~{Rubini}, G.~{Sindoni}, and G.~{Spandre}, ``{A very compact polarizer for
  an X-ray polarimeter calibration},'' {\em Presented at the Society of
  Photo-Optical Instrumentation Engineers (SPIE) Conference, San Diego (USA)
  26-30 August 2007}, 2007.

\bibitem{Muleri2007b}
F.~{Muleri}, P.~{Soffitta}, L.~{Baldini}, R.~{Bellazzini}, J.~{Bregeon},
  A.~{Brez}, E.~{Costa}, M.~{Frutti}, L.~{Latronico}, M.~{Minuti}, M.~B.
  {Negri}, N.~{Omodei}, M.~{Pinchera}, M.~{Razzano}, C.~{Sgro}, A.~{Rubini},
  and G.~{Spandre}, ``{Low energy polarization sensitivity of the Gas Pixel
  Detector},'' {\em submitted to Nuclear Instruments and Methods in Physics
  Research A} , July 2007.

\bibitem{Chandra1950}
S.~{Chandrasekhar}, {\em {Radiative transfer.}}, Oxford, Clarendon Press,
  1950., 1950.

\bibitem{SunTit1985}
R.~A. {Sunyaev} and L.~G. {Titarchuk}, ``{Comptonization of low-frequency
  radiation in accretion disks Angular distribution and polarization of hard
  radiation},'' {\em \aap}~{\bf 143}, pp.~374--388, Feb. 1985.

\bibitem{Connors1980}
P.~A. {Connors}, R.~F. {Stark}, and T.~{Piran}, ``{Polarization features of
  X-ray radiation emitted near black holes},'' {\em \apj}~{\bf 235},
  pp.~224--244, Jan. 1980.

\bibitem{Koyama1996}
K.~{Koyama}, K.~{Hamaguchi}, S.~{Ueno}, N.~{Kobayashi}, and E.~D. {Feigelson},
  ``{Discovery of Hard X-Rays from a Cluster of Protostars},'' {\em \pasj}~{\bf
  48}, pp.~L87--L92, Oct. 1996.

\bibitem{Revnivtsev2004}
M.~G. {Revnivtsev}, E.~M. {Churazov}, S.~Y. {Sazonov}, R.~A. {Sunyaev}, A.~A.
  {Lutovinov}, M.~R. {Gilfanov}, A.~A. {Vikhlinin}, P.~E. {Shtykovsky}, and
  M.~N. {Pavlinsky}, ``{Hard X-ray view of the past activity of Sgr A* in a
  natural Compton mirror},'' {\em \aap}~{\bf 425}, pp.~L49--L52, Oct. 2004.

\bibitem{Sunyaev1993}
R.~A. {Sunyaev}, M.~{Markevitch}, and M.~{Pavlinsky}, ``{The center of the
  Galaxy in the recent past - A view from GRANAT},'' {\em \apj}~{\bf 407},
  pp.~606--610, Apr. 1993.

\bibitem{Churazov2002}
E.~{Churazov}, R.~{Sunyaev}, and S.~{Sazonov}, ``{Polarization of X-ray
  emission from the Sgr B2 cloud},'' {\em \mnras}~{\bf 330}, pp.~817--820, Mar.
  2002.

\end{thebibliography}
\bibliographystyle{spiebib}   

\end{document}